\documentclass{article}
\usepackage{spconf,amsmath,graphicx, multirow, dblfloatfix}
\usepackage[hyphens]{url}

\title{Sound Event Detection Using Spatial Features and \\ Convolutional Recurrent Neural Network}
\name{Sharath Adavanne, Pasi Pertil\"{a}, Tuomas Virtanen\thanks{The research leading to these results has received funding from the European Research Council under the European Union’s H2020 Framework Programme through ERC Grant Agreement 637422 EVERYSOUND. The authors also wish to acknowledge CSC-IT Center for Science, Finland, for computational resources.}}
\address{Department of Signal Processing, Tampere University of Technology}

\begin{document}
%
\maketitle
\begin{abstract}
This paper proposes to use low-level spatial features extracted from multichannel audio for sound event detection. We extend the convolutional recurrent neural network to handle more than one type of these multichannel features by learning from each of them separately in the initial stages. We show that instead of concatenating the features of each channel into a single feature vector the network learns sound events in multichannel audio better when they are presented as separate layers of a volume. Using the proposed spatial features over monaural features on the same network gives an absolute F-score improvement of 6.1\% on the publicly available TUT-SED 2016 dataset and 2.7\% on the TUT-SED 2009 dataset that is fifteen times larger. 

\end{abstract}

 \vspace{-5pt}
\section{Introduction} \vspace{-5pt}
\label{sec:intro}

Sound event detection (SED) task involves recognizing the onset and offset of a sound event in an acoustic scene and further labeling the sound event. The world we live in offers a rich variety of sound events. For example, recognizing environmental sounds \cite{Karol2015}\cite{environmentalSED} will give an idea about the local biodiversity. Detecting sound events such as glass breaking and alarm detection can be used for surveillance \cite{surveillance_audio}\cite{surveillance}. Furthermore, the detected sound events can be used as a mid-level representation to help retrieval of content based query \cite{contentRetrieval}.

Traditionally SED systems have been using monaural audio. Temko et al. \cite{temko2007} proposed to use multichannel audio, and combined classification likelihoods across channels. While the multichannel audio was used, the actual potential of multichannel features was not exploited. Features like time difference of arrival (TDOA) and mel-band energies from the multichannel audio can potentially help the system differentiate the overlapping sound events. Similar multichannel features have been proposed in automatic speech recognition (ASR) \cite{multichannelASR} and source separation \cite{multichannelSS}. Just like humans have evolved to exploit the spatial data available at their ears (multichannel) to identify both isolated and polyphonic sound events \cite{iid_itd}, we can potentially train the SED systems to learn similar spatial information with multichannel data. Recently, such spatial features motivated by the binaural hearing of humans were proposed and shown to be promising for SED task in \cite{Adavanne2016_DCASE}. Although the features showed improvement over monoaural features, the dataset was too small (around one hour) to conclusively prove the superiority of binaural spatial features (referred as binaural features in future).

In this paper, we propose to use low-level features and compare it with using high-level features. For example, we compare using generalized cross-correlation with phase based weighting ($GCC$-$PHAT$) instead of the high-level TDOA feature which is extracted from $GCC$-$PHAT$, and show that the network learns powerful representation from just the low-level features. We show that arranging features from each channel as different layers of a multi-layered input volume enables the network to learn the sound events in multichannel audio better than a simple concatenation of the features. We propose to extend the convolutional recurrent neural network (CRNN) to handle more than one feature type and use a bi-directional LSTM. Finally, we evaluate the improvement of using binaural over monaural features on the 19 hours large TUT-SED 2016 dataset. 

We present the binaural features used for SED in Section 2, the extended CRNN architecture in Section 3, the experimental set-up and results on two different real-life datasets in Section 4 and our conclusions in Section 5.

 \vspace{-10pt}
\section{Binaural features for polyphonic SED} \vspace{-10pt}
\label{sec:feat}
Polyphonic SED is the task of recognizing overlapped sound events along with the isolated sound events. The proposed polyphonic SED system has two parts, feature extraction, and a neural network. The neural network described in Section \ref{sec:crnn} outputs a vector for every sound event class, where each entry in the vector indicates if the sound event was active or not. The feature extraction part extracts the following binaural features at a constant hop length of 20 ms.

\vspace{-10pt}
\subsection{Binaural mel-band energies} \vspace{-5pt}
\label{ssec:mel}
Sound sources which have different spatial locations have different intensities in the binaural channels. Furthermore, most overlapping sound events have different frequency spread in the spectrum. The combination of this intensity difference in different bands of frequencies can be exploited to differentiate overlapping sound events. This idea is motivated from the interaural intensity difference (IID) used by humans \cite{iid_itd}. 

Log mel-band energies (referred as $mel$ in future) extracted from both of the binaural channels using 40 mel-bands in 40 ms Hamming window are used as the features. A neural network which is capable of performing linear operations, which includes the difference, can learn to obtain the IID information from these channel-wise energies. By using the channel-wise energies instead of the multichannel energy difference directly, we allow the network to learn other potentially more informative features.

 \vspace{-10pt}
\subsection{Time difference of arrival vs cross-correlation} \vspace{-5pt}
\label{ssec:TDOA}
Based on how the sound sources are spatially located with respect to the binaural microphones, they might have different $TDOA$ values. Furthermore, sound events which are overlapping do not always have the same frequency spread in the spectrum. The combination of this $TDOA$ difference in different frequency bands can be exploited by a network to differentiate overlapping sound events. We implemented it by dividing the spectral frame into five mel-bands and calculating the $TDOA$ values in each of the bands. The $TDOA$ is estimated using the $GCC$-$PHAT$ ~\cite{GCC_PHAT}. The $GCC$-$PHAT$ for each mel-band $b$ is extracted separately: \vspace{-5pt}
\begin{equation}\vspace{-5pt}
 \label{eqn:1}
 R_b(\Delta_{12},t) = \sum_{k=0}^{N-1} H_b(k) \frac{  X_1(k,t) \cdot X_2^\ast(k,t)}{\arrowvert X_1(k,t)\arrowvert \arrowvert X_2(k,t)  \arrowvert} e^{ \frac{i 2\pi k \Delta_{12}}{N}}, \vspace{-5pt}
\end{equation} 
where, $X_1$ and $X_2$ are the FFT coefficients of the two binaural channels. $X_1(k,t)$ specifies the coefficient at time frame $t$ and $k$th frequency bin, of the total $N$ bins. ${H}_b(k)$ is the magnitude response of the $b$th band in $B$ mel-bands and $\Delta_{12} \in [-\tau_\textrm{max}, \tau_\textrm{max}]$, where $\tau_\textrm{max} = 30$ is the maximum sample delay for a sound wave to travel between binaural microphones. Finally, the peak magnitude for each mel-band and time frame is picked in the $GCC$-$PHAT$ by $\tau(b,t)= \underset{\Delta_{12}}{\textrm{argmax}}\left\{ |R_b(\Delta_{12},t)|\right\}$.

$TDOA$'s for each band are extracted using multi-resolution windows of 120 ms, 240 ms, and 480 ms to accommodate sound events of variable length. Five $TDOA$ values picked from five bands, for each of the three resolutions, results in 15 $TDOA$ values per time frame.

Neural networks have the potential to learn powerful representations from the raw data. We investigate this by using low-level $GCC$-$PHAT$ and comparing it with high-level $TDOA$ feature (which are picked from the $GCC$-$PHAT$). $GCC$-$PHAT$'s are extracted using Eq. \ref{eqn:1} with $B$ set to one. To have a factorizable feature length for max pooling, 60 $GCC$-$PHAT$ values are picked in -29 to +30 lag for each of the three multi-resolution (same as $TDOA$), amounting to 180 $GCC$-$PHAT$ values per time frame. By using $GCC$-$PHAT$ instead of $TDOA$, we take the data-oriented approach and get rid of empirical limitations and let the network learn the representation best suited for the problem.

 \vspace{-10pt}
\subsection{Dominant frequencies vs auto-correlation} \vspace{-5pt}
\label{ssec:pitch}

In \cite{Adavanne2016_DCASE}, it was shown that the three most dominant frequencies and their magnitudes (referred as $dom$-$freq$ in future) helped in the SED task. This was motivated by the idea that overlapping sound events do not always have the same dominant frequencies, and the network can learn to differentiate these overlapped events using the dominant frequencies. The $dom$-$freq$ values were picked from thresholded parabolically-interpolated STFT \cite{jos} in the 100 to 4000 Hz range from each of the binaural channels in frames of 40 ms. We continue to use this feature in this paper.


The pitch is a perceptual feature which human listeners have been using to recognize overlapping sound events \cite{pitchSED1990}. One of the prominent way to estimate pitch values are from the auto-correlation ($ACR$). In the presented work, $ACR$ is calculated on the binaural channels by time domain auto-correlation in 40 ms windows and choosing 400 correlation values in the range of 107.5 Hz to 4410 Hz. This was selected to be close to the  $dom$-$freq$ extraction range and the number of correlation values easily factorizable during max pooling.

 \vspace{-10pt}
\section{Convolutional Recurrent Neural Network} \vspace{-10pt}
\label{sec:crnn}
The best results to date in polyphonic SED was reported in \cite{emre_TASLP2016}, where an architecture exploiting the combined modeling capacities of a convolutional neural network (CNN), recurrent neural network (RNN) and fully connected (FC) layer termed as the convolutional recurrent neural network (CRNN) was proposed. We use this CRNN network and extend it for multichannel audio features.

Features from each channel of the multichannel features are layered one over the other to form a volume. More concretely, $M$ frames of a feature, each of length $L$, from two channels are layered into a $M\times L \times2$ volume. On slicing such a volume along a particular time frame, we get all the multichannel features corresponding to that time frame. The two-dimensional CNN's by design are built to learn on such volumes, i.e., it initially learns channel-wise filter weights, and further builds an activation map that is obtained as a combination of these channel-wise filter weights, which serves as the inter-channel information. This way we enable the CNN layers in the initial stages of the CRNN network to learn inter-channel information from multichannel features. We report the improvement in performance of using such a volume input over simple multichannel feature concatenation ($M\times 2L$) in Section \ref{ssec:results}.   

Separate volumes of each of the multichannel features are created. $T$ time frames of 40 $mel$ features from the two binaural channels are layered into one volume of size $T\times40\times2$. When using $dom$-$freq$, dominant frequencies and their magnitudes are treated as different features, and since their feature lengths are the same (3) we layer them in $T\times3\times4$. For $ACR$ we layer the 400 correlation values of each channel into a $T\times400\times2$ volume. Similarly, the three multi-resolution $TDOA$ features are layered to $T\times5\times3$ and the 60 values of $GCC$-$PHAT$ are layered to $T\times60\times3$. 

\begin{figure}
  \centering
  \includegraphics[width=\columnwidth]{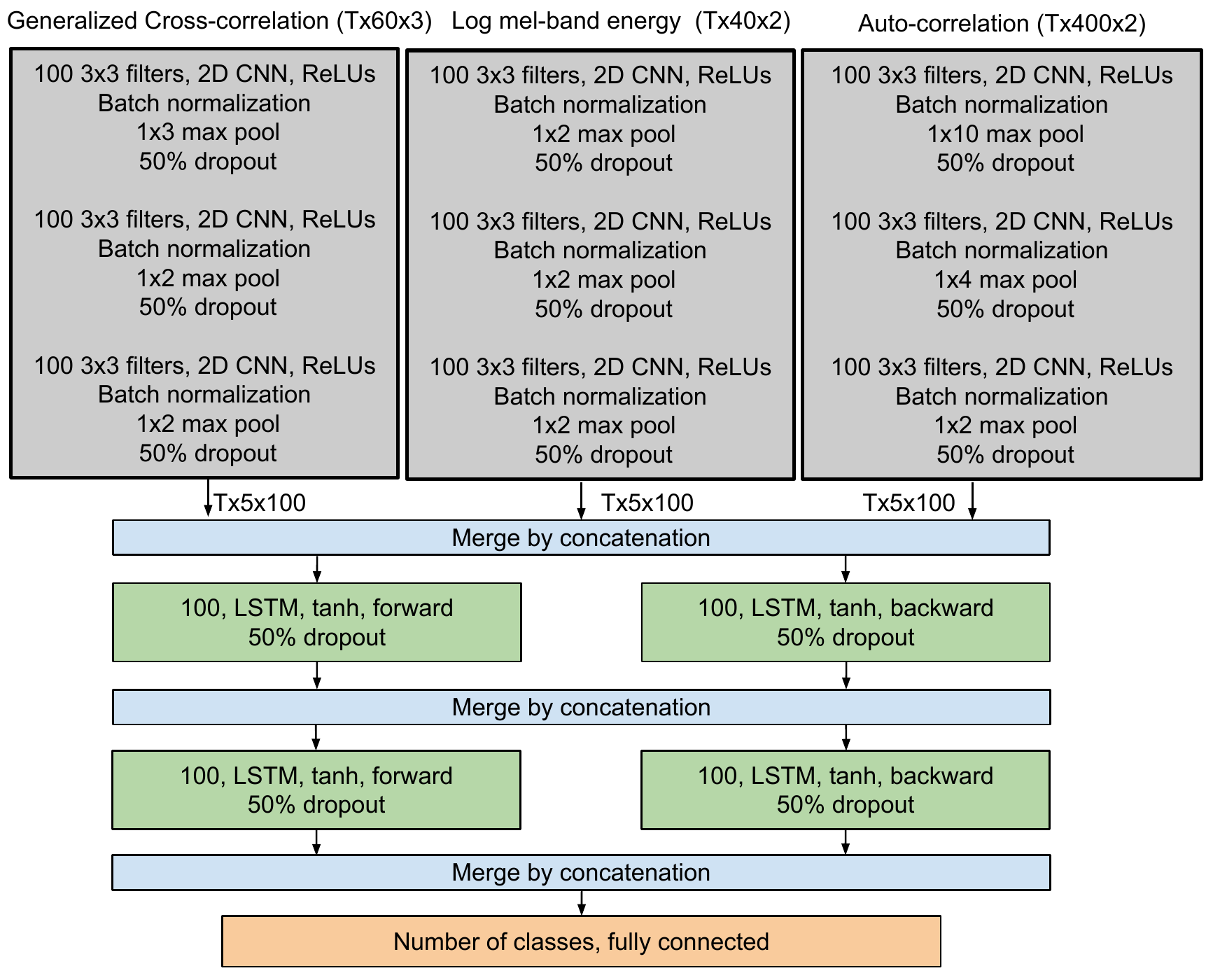}\vspace{-5pt}
  \caption{Convolutional bi-directional recurrent neural network (CBRNN) architecture for multichannel audio features}\vspace{-15pt}
  \label{fig:crnn}
\end{figure}

Separate CNN's are used to learn local shift-invariant features in each of these volumes as shown in Figure \ref{fig:crnn}. Since the dimensions of $mel$, $GCC$-$PHAT$, and $ACR$ are high, we use three CNN layers followed by max pooling to reduce the final feature map dimension to $T\times5\times100$. When using $TDOA$ and  $dom$-$freq$ features, a single 100-filter CNN layer is used without max pooling. To keep the time information intact for final sound event onset and offset detection, we do not apply max pooling in time ($T$) axis. Post CNNs, the feature maps are merged using concatenation and fed to two consecutive bi-directional long short term memory (LSTM). The output layer is a fully-connected time distributed layer which has as many units as the number of classes in the dataset. A sigmoid activation function is used at the output layer to allow several classes to be predicted as active simultaneously. We refer to this as the CBRNN system in future.

Batch normalization \cite{batchNorm} is used in all the CNN layers. A 50\% dropout \cite{Dropout} is utilized in all CNNs and LSTMs to avoid over-fitting of the network.  The combined architecture was trained by backpropagation through time \cite{bptt1990} using Adam optimizer \cite{adamKeras} and binary cross-entropy objective. Early stopping was used to reduce overfitting if the F-score (Section \ref{ssec:metrics}) did not change for 50 epochs. A sequence length of 100 frames (2 seconds) and a batch size of 32 was chosen after calibrating. At test time the sigmoid layer outputs are thresholded with a fixed value of 0.5. 

 \vspace{-5pt}
\section{Evaluation and Results} \vspace{-10pt}
\label{sec:typestyle}

\subsection{Datasets}\vspace{-5pt}
\label{ssec:data}
The proposed SED system is evaluated on two real-life datasets -TUT Sound Events 2009 (TUT-SED 2009) \cite{Heittola2010_EUSIPCO} and TUT Sound Events 2016 Development set (TUT-SED 2016)  \cite{Mesaros2016_EUSIPCO}. Both datasets have been recorded using in-ear microphones. TUT-SED 2009 has been used for SED in monaural context \cite{emre_TASLP2016}, but no previous work has reported using the binaural recordings on this dataset. TUT-SED 2016 was published as part of the DCASE 2016 challenge \cite{dcase2016task3web}, to allow public benchmarking. TUT-SED 2009 is fifteen times larger than TUT-SED 2016, by showing considerable improvement on TUT-SED 2009 we can conclusively say the proposed system is learning and exploiting spatial information. 

All the work proposed in this paper is done in a context-independent manner, i.e., we train a single system to learn sound event classes across contexts.

The first dataset - TUT-SED 2009 consists of 103 binaural recordings from 10 different contexts (listed in Table \ref{Table:3}). Each context consists of 8 to 14 recordings which vary from 10 to 30 minutes, amounting to an overall length of 1133 minutes. The recordings have been manually annotated, and the annotated events have been grouped into 61 event classes \cite{Heittola2010_EUSIPCO}. Each context has 9-16 event classes, while some events occur in multiple contexts, some are context specific. The dataset defines five-folds for training, validation, and testing. 

The second dataset - TUT-SED 2016 consists of 22 binaural recordings for two contexts - home and residential area, amounting to 78 minutes. The home context has ten recordings with 11 sound event classes, and the residential area has 12 recordings with seven sound event classes \cite{Mesaros2016_EUSIPCO}. The dataset defines four-folds for training and testing. We use 20\% of the training data for validation, and the same validation is used for all our evaluations.
\begin{table*}[!b]\centering\vspace{-10pt}
\setcounter{table}{1}
\resizebox{\textwidth}{!}{\begin{tabular}{l|ccccccc|ccc}
\multirow{2}{*}{Feature combination} & \multicolumn{7}{c}{Indoor} & \multicolumn{3}{c}{Outdoor} \\
 & Basketball & Bus & Hallway & Office & Car & Restaurant & Shop & Beach & Street & Track and Field \\
\hline
$mel$-$monaural$  & 79.7 & 52.6 & 59.1 & 81.8 & 78.2 & 80.7 & 62.4 & 56.5 & 60.3 & 70.1 \\
\hline
$mel$ & 82.2 & 56.5 & 66.6 & 83.3 & 81.5 & 83.1 & 63.3 & 59.5 & 66.0 & 70.2 \\
\hline
$mel$ + $TDOA$ & 82.8 & 58.7 & 66.0 & 80.8 & 79.2 & 81.2 & 64.7 & \textbf{60.9} & \textbf{66.9} & 68.8 \\
$mel$ + $GCC$-$PHAT$ & 81.9 & 58.9 & 65.3 & 80.0 & 81.2 & 81.3 & 65.3 & 60.4 & 66.3 & \textbf{72.6} \\
\hline
$mel$ +  $dom$-$freq$ & \textbf{83.7} & \textbf{60.5} & \textbf{67.8} & \textbf{84.6} & 80.8 & 81.8 & 64.6 & 60.7 & 66.6 & 67.6 \\
$mel$ + $ACR$ & 82.9 & 58.6 & 63.8 & 83.6 & \textbf{83.4} & \textbf{82.3} & \textbf{65.5} & 60.4 & 65.8 & 69.0 \\
\hline
$mel$ + $TDOA$ +  $dom$-$freq$ & 83.0 & 59.4 & 67.5 & 83.9 & 78.6 & 79.9 & 65.1 & 60.5 & 65.0 & 70.0 \\
$mel$ + $GCC$-$PHAT$ + $ACR$ & 82.8 & 59.1 & 66.8 & 82.2 & 79.4 & 80.4 & 64.8 & 60.4 & 66.1 & 68.5
\end{tabular}}\vspace{-5pt}
\caption{Context wise F-scores for TUT-SED 2009 dataset.} \vspace{-10pt}
\label{Table:3}
\end{table*}

\vspace{-10pt}
\subsection{Metrics}\vspace{-5pt}
\label{ssec:metrics}
The SED system output is evaluated with the reference in fixed length intervals, also called as segment-based evaluation \cite{metrics}. For each segment $k$, the following are calculated (i) true positive ($TP(k)$): total number of events active in both reference and system output segment. (ii) False positive ($FP(k)$): total number of events active in system output segment but not in reference. (iii) False negative ($FN(k)$): total number of events active in reference segment but not in system output. The first metric, F-score is then calculated as,\vspace{-5pt}
\begin{equation}
F = \frac{2 \cdot \sum_{k=1}^{K} TP(k)}{2 \cdot \sum_{k=1}^{K}TP(k)+ \sum_{k=1}^{K}FP(k)+ \sum_{k=1}^{K}FN(k)}
\end{equation}

The second metric, error rate (ER) evaluates the system output based on the number of insertions (I), deletions (D) and substitutions (S). \vspace{-5pt}
\begin{align}\vspace{-10pt}
ER = \frac{\sum_{k=1}^{K} S(k) + \sum_{k=1}^{K} D(k) + \sum_{k=1}^{K} I(k)}{\sum_{k=1}^{K} N(k)}
\end{align}
Where $N(k)$ is the number of sound events marked as active in the reference segment $k$, and\vspace{-5pt}
\begin{align}\vspace{-10pt}
S(k) = \min(FN(k), FP(k)) \\
D(k) = \max(0, FN(k)-FP(k)) \\
I(k) = \max(0, FP(k)-FN(k)) 
\end{align}
We use a segment length of one second for ER and F-score estimation. The evaluation metrics are calculated for each context separately and averaged result is presented.

 \vspace{-10pt}
\subsection{Baseline} \vspace{-5pt}
The proposed CBRNN architecture with binaural features is compared with the state of the art monaural SED system introduced in \cite{emre_TASLP2016}. The system used 40 monaural log mel-band energies ($mel$-$monaural$) as features. The network had three CNN's each of 96 filters, followed by max pooling in frequency axis reducing the dimension to one. The feature map from CNN was then fed to three LSTMs with 256 units each. The output was a fully-connected layer with units equal to the number of classes in the dataset. 

\vspace{-10pt}
\subsection{Results} \vspace{-5pt}
\label{ssec:results}

Table \ref{Table:2} shows the metrics for multi-layered input of the binaural log mel-band energy features ($mel$) and concatenating it ($mel$-$concat$) for TUT-SED 2009 dataset. Using a multi-layered input is seen to perform relatively better than a simple concatenation. Similar improvement was observed using multi-layered input of $TDOA$,  $dom$-$freq$, $GCC$-$PHAT$ and $ACR$ (not tabulated).

From Table \ref{Table:2} we see that using binaural features improves both the ER and F-scores over monaural features ($mel$-$monaural$) across datasets. While the  $dom$-$freq$ and $mel$ feature combination gave the best performance in TUT-SED 2009, $TDOA$ and $mel$ performed the best for TUT-SED 2016. In numbers, using binaural over monaural features on the same network gives an absolute F-score improvement of 2.7\% for TUT-SED 2009 and 6.1\% for TUT-SED 2016. By showing this improvement on a larger dataset like TUT-SED 2009, we can more confidently say that the network is truly learning the binaural information.

From the metrics in Table \ref{Table:2} and \ref{Table:3} we see that the performance of using $GCC$-$PHAT$ instead of $TDOA$ or $ACR$ instead of $dom$-$freq$, is comparable. This is a significant result, showing that the network can learn equivalent information of powerful high-level features from just the low-level features. Thereby making the features dataset independent and relieving the tuning of parameters like the number of $dom$-$freq$ and $TDOA$ values.

Most of the sound event classes were seen to be recognized better with the binaural features. Since we cannot present all the 79 classes of the two datasets in this paper, we show the context based F-scores for TUT-SED 2009 dataset in Table \ref{Table:3}. A general observation is that the $dom$-$freq$ / $ACR$ and $mel$ are useful for indoor and sound intense environment (bus, hallway, office, and basketball), while $TDOA$ / $GCC$-$PHAT$ and $mel$ are seen to help in outdoor contexts (beach and street). This also explains why  $dom$-$freq$ and $mel$ gave better results for TUT-SED 2009. While TUT-SED 2016 had one each of indoor and outdoor contexts, TUT-SED 2009 had more indoor contexts than outdoor.

The proposed CBRNN architecture using the same $mel$-$monaural$ feature used in CRNN-baseline achieved an F-score of 68.0\% for TUT-SED 2009 and 29.7\% for TUT-SED 2016 (Table \ref{Table:2}). The difference in the scores with respect to CRNN-baseline can be associated with using a higher dimensional input to LSTM's in the proposed CBRNN.

\begin{table}[!t]  
\setcounter{table}{0}
\centering
\resizebox{0.48\textwidth}{!}{\begin{tabular}{l|cc|cc} 
\multicolumn{1}{c}{\multirow{2}{*}{Feature combination}} & \multicolumn{2}{c}{TUT-SED 2009} & \multicolumn{2}{c}{TUT-SED 2016} \\
\multicolumn{1}{c}{} & ER & F & ER & F \\
\hline
CRNN baseline \cite{emre_TASLP2016} & 0.49 & 68.8 & \textbf{0.93} & 31.3 \\
$mel$-$monaural$ & 0.49 & 68.0 & 1.03 & 29.7 \\
\hline
$mel$-$concat$ & 0.44 & 70.3 &  &  \\
$mel$ & \textbf{0.43} & 71.1 & 0.99 & 32.3 \\
\hline
$mel$ + $TDOA$ & 0.45 & 70.9 & 0.95 & \textbf{35.8} \\
$mel$ + $GCC$-$PHAT$ & 0.44 & 71.1 & 0.95 & 34.6 \\
\hline
$mel$ +  $dom$-$freq$ & \textbf{0.43} & \textbf{71.7} & 0.98 & 32.8 \\
$mel$ + $ACR$ & 0.44 & 71.2 & 0.98 & 33.8 \\
\hline
$mel$ + $TDOA$ +  $dom$-$freq$ & 0.44 & 71.0 & 1.01 & 33.3 \\
$mel$ + $GCC$-$PHAT$ + $ACR$ & 0.45 & 70.9 & 0.99 & 33.6
\end{tabular}}  \vspace{-10pt}
\caption{Error rate (ER) and F-score achieved using binaural features and CBRNN on TUT-SED 2009 and 2016 datasets.}  \vspace{-10pt}
\label{Table:2}
\end{table}  

 \vspace{-10pt}
\section{Conclusion} \vspace{-10pt}
In this paper, we extended convolutional recurrent neural networks to handle multiple feature classes and process feature-maps using bi-directional LSTM's. A multi-layered input of multichannel features which enables the network to learn sound events in a multichannel audio better was proposed. Low-level features were used in place of high-level features, and the network was shown to learn high-level equivalent information from simple low-level features. The performance of the system was evaluated on two datasets - a larger dataset for proving that the binaural features truly help in improving the sound event detection, and a public dataset, to allow other researchers to benchmark. The proposed network using binaural spatial features was shown to recognize sound events better than using just the monaural features.

\bibliographystyle{IEEEtran}
\bibliography{refs,strings}

\end{document}